\documentclass[twocolumn,apj,dvipdfm]{emulateapj}
\usepackage{natbib}
\usepackage{color}
\usepackage{graphicx}
\usepackage{hhline}
\usepackage{CJK}

\def\mbf#1{\mbox{\boldmath ${#1}$}}
\def\Alfven{Alfv\'{e}n~}
\def\Alfvenic{Alfv\'{e}nic~}

\begin{document}
\begin{CJK*}{UTF8}{gbsn}
\title{Driving Disk Winds and Heating up Hot Coronae by MRI Turbulence}

\author{Yuki Io (井尾 勇き)\altaffilmark{1,2}  \& Takeru K. Suzuki (すず木 建)\altaffilmark{1}}
\email{stakeru@nagoya-u.jp}
\altaffiltext{1}{Department of Physics, Nagoya University,
Furo-cho, Chikusa-ku, Nagoya, Aichi 464-8602, Japan}
\altaffiltext{2}{Nihon Unisys, Ltd., Toyosu, Koto-ku, Tokyo, 
135-8560, Japan}

\begin{abstract}
We investigate the formation of hot coronae and vertical outflows in accretion 
disks by magneto-rotational turbulence.
We perform local three-dimensional (3D) MHD simulations with the vertical 
stratification by explicitly solving an energy equation with various 
effective ratios of specific heats, $\gamma$. 
Initially imposed weak vertical magnetic fields are effectively 
amplified by magnetorotational instability (MRI) and winding due to the 
differential rotation.
In the isothermal case ($\gamma=1$), the disk winds are mainly driven by 
the Poynting flux associated with the MHD turbulence and show quasi-periodic 
intermittency.
On the other hand, in the non-isothermal cases with $\gamma \ge 1.1$, 
the regions above 1-2 scale heights from the midplane are effectively 
heated up to form coronae with the temperature of $\sim$ 50 times of the 
initial value, which are connected to the cooler midplane region through 
the pressure-balanced transition regions. 
As a result, the disk winds are mainly driven by the gas pressure with 
exhibiting more time-steady nature, although the nondimensional
time-averaged mass loss rates are similar to that of the isothermal case. 
Sound-like waves are confined in the cool midplane region in these cases, 
and the amplitude of the density fluctuations is larger than that of the 
isothermal case. 

\end{abstract}

\section{Introduction}


In accretion disks magnetohydrodynamical (MHD) turbulence is believed to work 
as effective viscosity and play an essential role in the outward transport 
of the angular momentum and the radial motion of 
the gas \citep[e.g.][]{lp74,bh98}. 
Magnetorotational instability \citep[MRI hereafter][]{vel59,cha60,bh91} is 
a promising source of such turbulent viscosity in accretion disks. 

Properties of the MRI in accretion disks have been widely studied. 
MHD simulations in the local shearing coordinates \citep{gl65} 
have been extensively and intensively performed without vertical density 
stratification 
\citep[e.g.,][]{hgb95,mt95,san04}, 
and with the vertical stratification due to 
the gravity by a central object 
\citep[e.g.,][]{sto96,ms00,dav10,skh10,oh11,sai13}.

Accretion disks threaded with global magnetic fields are supposed to drive 
disk winds, which was first suggested by \citet{bp82} as an origin of 
jets from black hole accretion disks. This picture of 
magnetocentrifugally driven disk winds was extended to protostellar jets 
\citep{pn83}.  
Outflows and jets from various types of accretion disks have 
been observed, {\it e.g.}, from young stars \citep{oh97,cof04}, and 
from active galactic nuclei \citep{tom10}.

These two pictures -- MRI-driven turbulence in accretion disks and 
magnetically driven winds from disk surfaces involving coherent field lines -- 
are supposed to have a close link. 
For instance, turbulence in disks could drive the vertical motions 
of the gas by the MHD turbulent pressure. 
Based on these considerations, \citet{suz09} and \citet{suz10} recently 
proposed the 
onset of vertical outflows from MRl-turbulent accretion disks. They performed 
MHD simulations in local shearing boxes with the vertical stratification and 
net vertical magnetic fields by taking a special care of the outgoing 
boundary conditions \citep{suz06} at the vertical surfaces of the simulation 
boxes. They found that the
disk winds are driven from the upper and lower boundaries by the Poynting flux 
associated with the MHD turbulence. 
This process can contribute to the mass loading to the basal 
regions of the global disk winds introduced above. 
Such vertical outflows are also observed in 3D global simulations 
\citep{flo11,suz13}.

Later on, such MRI-turbulent driven disk winds have been further studied 
from various aspects. 
\citet{bs13a} examined properties of the disk winds 
with stronger net vertical magnetic fields. The connection between the upflows 
in local simulations and the global disk winds/jets is investigated 
\citep{les13,bs13a}.
\citet[][see also Suzuki \& Inutsuka 2010]{fro13} pointed out the mass flux
of the disk winds depends on the simulation box size, which shows 
that we need great cares to handle disk winds by local shearing boxes. 

Although there are limitations in the shearing box approximation, 
it is still useful technic to study the basic properties of MRI-turbulent 
driven vertical outflows. 
Most of previous simulations adopt an isothermal equation 
of state to mainly focus on the dynamics of the disk winds, apart from 
a limited number of works that considered detailed heating and cooling 
processes for black hole accretion disks \citep{tur04,hir06} and 
protoplanetary disks \citep{ht11}. 
So far there has been no systematic studies done for the MRI-turbulent driven 
winds with different ratios of specific heats, $\gamma$ even within a framework 
of the shearing box approximation, which is the main focus of the present 
paper. 


\section{Simulation Setups}
In \citet{suz09} \& \citet{suz10} we performed 3D MHD simulations 
in local stratified shearing boxes \citep{sto96} by solving the ideal MHD 
equations with an isothermal equation of state.  
In this paper, we extend these works by explicitly 
solving an energy equation in a Lagrangian form, 
$$
\hspace{-1cm}\rho \frac{d}{dt}\left[e+\frac{v^2}{2}+\frac{B^2}{8\pi\rho} 
+\frac{\Omega^2}{2}(z^2-3x^2)\right] 
$$
\begin{equation}
= \mbf{\nabla}\cdot\left[-\left(p+\frac{B^2}{8\pi}\right)\mbf{v}
+\frac{\mbf{B}}{4\pi}(\mbf{B\cdot v})\right], 
\label{eq:eng}
\end{equation}  
where $\Omega$ is the Keplerian rotation frequency,  
$e$ is the specific energy per mass which is related to the gas pressure, $p$, 
the density, $\rho$, and the effective ratio of specific heats, $\gamma$, as 
\begin{equation}
e = \frac{1}{\gamma-1}\frac{p}{\rho}, 
\end{equation} 
and the other variables in Equation (\ref{eq:eng}) have the conventional 
meanings.
The terms involving $\Omega^2$ in Equation (\ref{eq:eng}) originate from 
the central object; $\Omega^2z^2/2$ is the potential due to the vertical 
component of the gravity and $-3\Omega^2x^2/2$ denotes the tidal potential.

In the energy equation above we do not explicitly consider external cooling 
and heating processes, {\it e.g.,} radiation cooling/heating, thermal 
conduction, and {\it etc}.  Instead, we study their effect 
phenomenologically by assuming different but spatially 
uniform $\gamma$ from 1 to $5/3$ in different cases. 
In our simulations, the gas is heated up mainly by the dissipation of 
magnetic energy, which we discuss later in this section. 
Taking $\gamma=1$ (isothermal condition) indicates 
that we (implicitly) assume that the temperature is kept constant by an 
unspecified cooling that can balance the magnetic heating. On the other hand, 
larger $\gamma$ corresponds to suppressing cooling; $\gamma=7/5$ 
and $5/3$ correspond to the adiabatic conditions for diatomic 
and monoatomic gases, respectively.

Apart from explicitly solving the energy equation, we basically follow 
\citet{suz09} when performing the numerical simulations. 
We adopt a second-order Godunov-CMoCCT scheme \citep{san99}, in which we 
solve the nonlinear Riemann problems with the magnetic pressure at the 
cell boundaries for the compressive waves and adopt the consistent method 
of characteristics (CMoC) for the evolution of magnetic fields \citep{cl96}. 
At the top and bottom $z$ boundaries, we prescribe the outgoing boundary 
condition by using the seven MHD characteristics \citep{suz06}.

We fix the ratio of the box sizes of the $x$, $y$, and $z$ axes 
to 1:4:8, which are respectively resolved by 32, 64, and 256 spatially uniform 
grids. Each cell is elongated in the $y$ direction with $\Delta y$ being 
twice as large as $\Delta x$ and $\Delta z$. 
The lengths in the simulations are normalized by the initial pressure 
scale heights, $H_0$, as 
\begin{equation}
H_0 = \frac{\sqrt{2}c_{\rm s,0}}{\Omega} \equiv \frac{\sqrt{2T_0}}{\Omega},
\end{equation} 
where $c_{\rm s,0}$ is the initial sound speed, and we also use the initial 
temperature, $T_0$, which has the dimension of $v^2$.  Hereafter subscripts 
`0' represent the initial state. 
In this paper, $c_{\rm s}$ stands for an ``isothermal'' sound speed, 
$=\sqrt{p/\rho}=\sqrt{T}$; the usual sound speed, 
$\sqrt{(\partial p/\partial \rho)_{s}} =\sqrt{\gamma p/\rho}=\sqrt{\gamma T}$, 
is expressed as $\sqrt{\gamma}c_{\rm s}$.
We initially set up the hydrostatic density structure with the constant 
temperature, $T=T_0$, 
\begin{equation}
\rho = \max\left(\rho_{\rm mid,0} \exp\left(-\frac{z^2}{H_0^2}\right),
10^{-9}\rho_{\rm mid,0}\right), 
\label{eq:initrho}
\end{equation}
where $\rho_{\rm mid,0}$ is the initial density at the midplane.
In order to perform the simulations stably, we adopt a floor value, 
$\rho_{\rm fl} = 10^{-9}\rho_{\rm mid,0}$ throughout the simulations.
In the non-isothermal cases, the simulation boxes are larger than that of 
the isothermal case to treat the extended coronae as shown later. 
In the cases with $\gamma\ge 1.03$, the initial densities near the surface 
regions, $|z|>4.55H_0$, are smaller than $\rho_{\rm fl}$, hence $\rho_{\rm fl}$ 
is used for these regions. 
In the simulations, we use the unit of $\Omega=1$, $H_0=1$, and 
$\rho_{\rm mid,0}=1$; accordingly, $T_0 = 1/2$ and $c_{\rm s,0}=1/\sqrt{2}$. 
We initially impose the weak net vertical magnetic field, $B_{z,0}$, with the 
plasma $\beta$ value, $8\pi p_{\rm mid,0}/B_{z,0}^2=10^6$ at the midplane.
We start the simulations with giving small random perturbations
with 0.5\% of $c_{\rm s,0}$
as seeds of the MRI. The simulations are kept running until 200 rotations 
($t\Omega/2\pi=200$). 

In the simulations with $\gamma>1$, the gas is heated by the dissipation 
of the magnetic energy, where the heating is done by the numerical effect in 
the sub-grid scale since we do not explicitly take into account resistivity.
Physically, we expect that the heating is due to cascading turbulence 
at small scales.
As a disk is heated up, the sound speed increases, 
and accordingly the pressure scale height also increases. 
Therefore, to study the vertical structure, the $z$ coordinate in units of 
the scale height, 
\begin{equation}
\int_{0}^{z}\frac{dz}{\langle h(z)\rangle}_{x,y} \equiv 
\int_{0}^{z}\frac{dz\Omega}{\sqrt{2}\langle c_{\rm s}\rangle_{x,y}(z)},
\end{equation}
is a key quantity, where the subscripts, $x,y$, of $\langle \; \rangle$ 
indicate the average over a horizontal ($x-y$) plane. 
While $\int_{0}^{z}\frac{dz}{\langle h(z)\rangle_{x,y}}=z/H_0$ 
for the isothermal situation, generally 
$|\int_{0}^{z}\frac{dz}{\langle h(z)\rangle_{x,y}}|<|z/H_0|$ 
as a result of the heating for $\gamma > 1$. 

\begin{deluxetable}{ccccc}[h]
\tabletypesize{\footnotesize}
\tablecaption{Simulation runs with different ratios of specific heats, 
$\gamma$. \label{tab:bsz}}
\tablehead{\colhead{$\gamma$} & \colhead{$(X,Y,Z)/H_0$} 
& \colhead{$\int_{z_{\rm bot}}^{z_{\rm top}}{dz}{\langle h(z)\rangle}$} 
& \colhead{$\Delta t_{\rm ave}$(rot.)} & \colhead{$M_{\rm f}/M_0$} } 
\startdata
1.0 & 0.9,3.6,7.2 & 7.2 & 140--200 & 0.907\\
\hline
1.01 & 1.0,4.0,8.0 & 7.1 & 120--200 & 0.895\\
\hline
1.03 & 1.35,5.4,10.8 & 7.1 & 100--200 & 0.896\\
\hline
1.1 & 4.0,16.0,32.0 & 7.3 & 150--200 & 0.951\\
\hline
1.2 & 4.0,16.0,32.0 & 7.1 & 130--200 & 0.906\\
\hline
1.3 & 4.0,16.0,32.0 & 7.3 & 130--200 & 0.937\\
\hline
1.4 & 4.0,16.0,32.0 & 7.4 & 120--200 & 0.934\\
\hline
5/3 & 4.0,16.0,32.0 & 7.2 & 120--200 & 0.890 
\enddata
\tablecomments{The initial box size ({\it 2nd column}) and the final 
vertical box size measured in the final scale height ({\it 3rd column}; 
Equation \ref{eq:fvsz}) averaged over $\Delta t_{\rm ave}$ are compared. 
$X$ ,$Y$, and $Z (=z_{\rm top}-z_{\rm bot})$ 
in the 2nd column are the sizes of $x$, $y$, and $z$ components of 
the simulation box. The 4th column shows the final mass, $M_{\rm f}$, 
normalized by the initial mass, $M_{0}$, remained in the box.}
\end{deluxetable}

The mass flux of the disk winds depends on the vertical box size in units of 
the scale height \citep{suz10,fro13}.  
In order to compare the properties of the disk winds with different $\gamma$, 
it is desirable to adopt the same vertical box size in units of the final 
scale height after the steady-state conditions are achieved, 
\begin{equation}
\int_{z_{\rm bot}}^{z_{\rm top}}\frac{dz}{\langle h(z)\rangle_{\Delta t_{\rm ave},x,y}},
\label{eq:fvsz}
\end{equation}
where $z_{\rm bot}$ and $z_{\rm top}$ indicate the locations of the bottom 
and top boundaries of the simulation box, and we take the average over 
$\Delta t_{\rm ave}$ (Table \ref{tab:bsz}) after 
the magnetic fields are amplified to the saturated state, which 
we describe later in this section. We cannot estimate $h(z)$ in advance, hence 
we perform simulations with different box sizes and pick up one that gives 
the desirable value for each $\gamma$. 
For the isothermal case ($\gamma=1$), we fix the vertical box size 
$(z_{\rm top}-z_{\rm bot})=7.2H_0$, indicating $0.9H_0$ and $3.6H_0$ for the 
box sizes in the $x$ and $y$ directions. 
For the other cases with $\gamma>1$, we tune
the initial box sizes to give the final vertical box sizes after the 
saturation in a range of $7.1<\int_{z_{\rm bot}}^{z_{\rm top}}dz/\langle h(z)
\rangle_{\Delta t_{\rm ave},x,y}<7.4$. 

\begin{figure}[h]
\begin{center}
\includegraphics[width=0.4\textwidth]{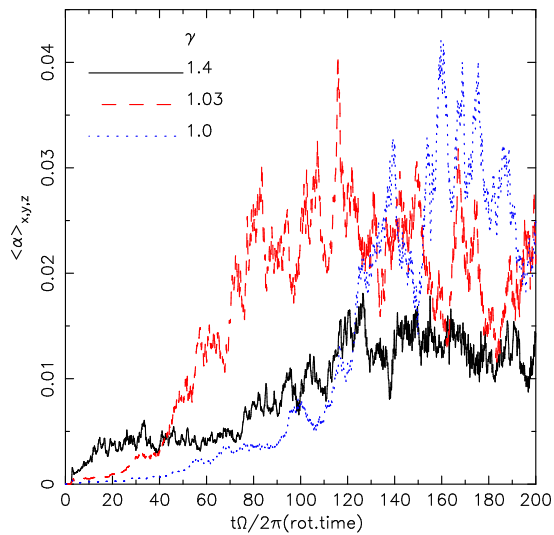}
\end{center}
\caption{Evolution of the box-averaged $\langle \alpha \rangle_{x,y,z}$ 
of the three cases with $\gamma=1.4$ ({\it black solid}), 1.03 ({\it red 
dashed}), and 1.0 ({\it blue dotted}). 
The vertical axis, $t\Omega/2\pi$, indicates time in units of one rotation.}
\label{fig:totalp}
\end{figure}

\begin{figure}[h]
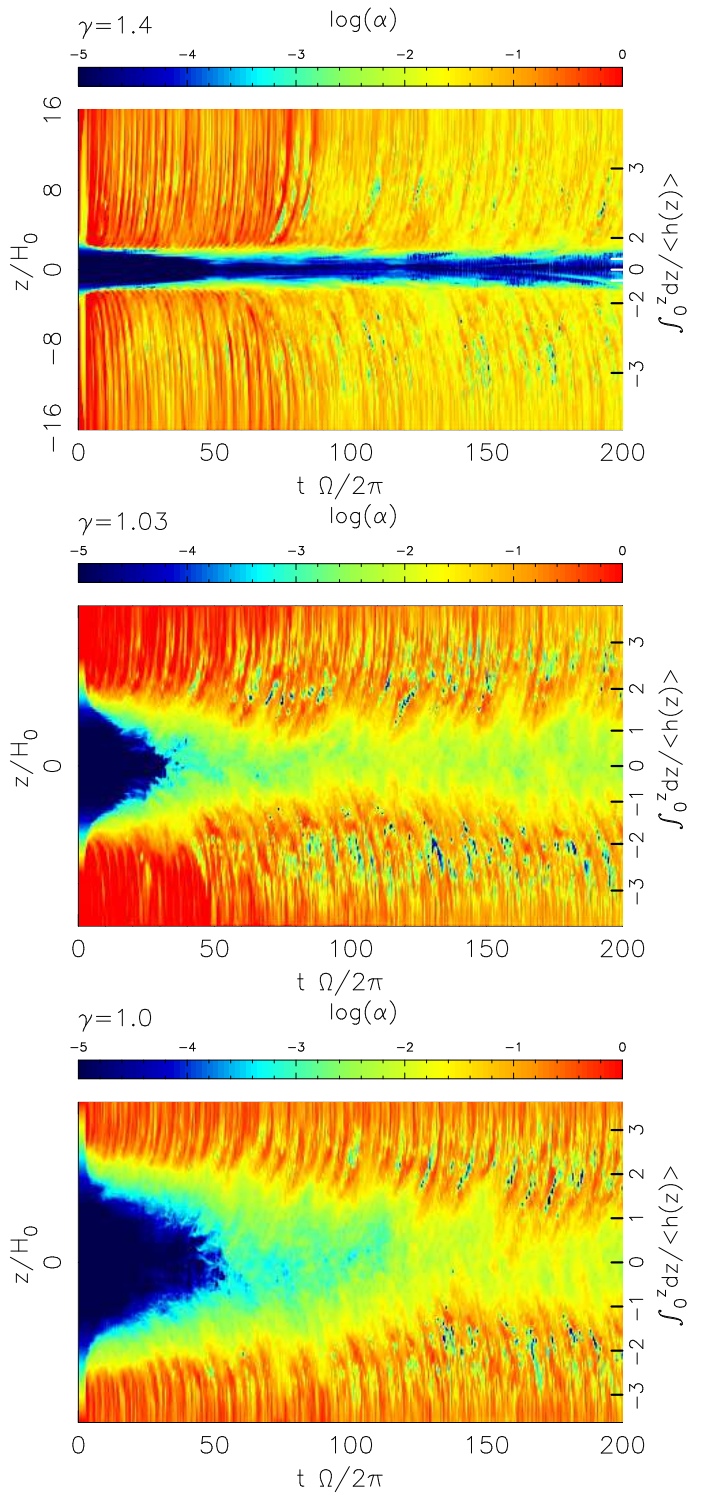

\begin{center}
\includegraphics[width=0.5\textwidth]{t-z_gamma14_alpha1.ps}\\
\includegraphics[width=0.5\textwidth]{t-z_gamma103_alpha1.ps}\\
\includegraphics[width=0.5\textwidth]{t-z_gamma10_alpha1.ps}
\end{center}
\caption{Time-distance diagrams of $\langle\alpha\rangle_{x,y}$ 
of the cases with $\gamma=1.4$ ({\it top}), 1.03 ({\it middle})
and 1.0 ({\it lower}). The horizontal axis ($t\Omega /2\pi$) 
denotes time in units of rotation. The vertical axis shows $z$; on the left 
shown is distance in units of the initial scale height, $z/H_0$, and 
on the right shown is distance in units of the final scale height, 
$\int_{0}^{z}dz/\langle h(z)\rangle_{\Delta t_{\rm ave},x,y}$. }
\label{fig:tzalpha}
\end{figure}

In order to check the saturation of the magnetic fields, we monitor 
the box-averaged $\alpha$ values \citep{ss73}, which is the sum of the
Reynolds and Maxwell stresses normalized by the gas pressure, 
\begin{equation}
\langle \alpha \rangle_{x,y,z} = \frac{\int_{z_{\rm bot}}^{z_{\rm top}}dz \langle
\rho v_x \delta v_y - B_x B_y/4\pi \rangle_{x,y}}{\int_{z_{\rm bot}}^{z_{\rm top}}dz 
\langle p\rangle_{x,y}}, 
\end{equation}
where $\delta v_y = v_y-v_{y,0}$ is the difference of toroidal velocity from 
the Keplerian shear flow, 
$v_{y,0}=-\frac{3}{2}\Omega x$.
In Figure \ref{fig:totalp} we show the results of the 
cases with $\gamma = 1.4$ (black solid line), 1.03 (red dashed line), 
and 1.0 (blue dotted line). 
We also display the time evolution of the vertical structure of 
horizontally averaged $\langle \alpha \rangle_{x,y}=\langle\rho v_x 
\delta v_y - B_x B_y/4\pi \rangle_{x,y}/ \langle p\rangle_{x,y}$ 
in $t$--$z$ diagrams in Figure \ref{fig:tzalpha} to see more details.
In the case with $\gamma = 1.4$, the 
saturation is observed after $t\Omega /2\pi
\gtrsim 120$, and we use $\Delta t_{\rm ave}=120-200$ rotations 
(1 rotation$=2\pi/\Omega$) 
to estimate the vertical box size in the final scale height 
(Equation \ref{eq:fvsz}).
Although in the case with $\gamma=1.03$ the box averaged 
$\langle \alpha \rangle_{x,y,z}$ almost saturates after $t\Omega /2\pi 
\gtrsim 80$ (Figure \ref{fig:totalp}), we use a more conservative  
$\Delta t_{\rm ave}=100-200$, because $\langle \alpha\rangle_{\Delta t_{\rm ave},x,y}$ 
at the midplane is growing in $t\Omega /2\pi \lesssim 90$ (middle panel 
of Figure \ref{fig:tzalpha}).
In the isothermal case ($\gamma=1.0$), we use $\Delta t_{\rm ave}=140-200$ 
rotations, because $\langle \alpha \rangle_{x,y,z}$ becomes saturated later 
than in the other two cases. 
On the right of each panel of Figure \ref{fig:tzalpha}, 
we show $z$ in units of the final scale height, $\int_0^zdx/\langle h(z)
\rangle_{\Delta t_{\rm ave},x,y}$. The interval between the ticks with $\Delta z=1$ 
varies with height in the non-isothermal cases because the temperature 
is higher near the surfaces as shown in $\S$ \ref{sec:res}.

The initial and final box sizes and $\Delta t_{\rm ave}$'s are summarized 
in Table \ref{tab:bsz}. 
For the cases with $\gamma\ge 1.1$, the vertical box size measured in 
the final scale height shrinks more than a factor of 4 because of the heating, 
which we examine in \S \ref{sec:res}.
The derived $\Delta t_{\rm ave}$ is also used to examine the time-averaged 
vertical structures of each case in \S \ref{sec:res}.  

The horizontal box sizes measured in the scale height of the non-isothermal 
cases vary with height because of the variation of the temperature.  
Because of the higher temperature near the surfaces (\S \ref{sec:res}), 
the horizontal box sizes in units of the final scale height, 
$\int_{x_{\rm min}}^{x_{\rm max}} dx/\langle h(z)\rangle_{\Delta t_{\rm ave},x,y}<0.9$ 
and $\int_{y_{\rm min}}^{y_{\rm max}} dy/\langle h(z)\rangle_{\Delta t_{\rm ave},x,y}<3.6$, 
are smaller in the surface regions. 
of the non-isothermal cases. 
Although these horizontal sizes are 
insufficient to quantitatively discuss the saturation of the magnetic fields 
\citep[e.g.,][]{gu09}, we anticipate that it is still meaningful to compare 
the vertical disk structures with different $\gamma$. 

\section{Results}
\label{sec:res}
\begin{figure}[h]
\begin{center}
\includegraphics[width=0.5\textwidth]{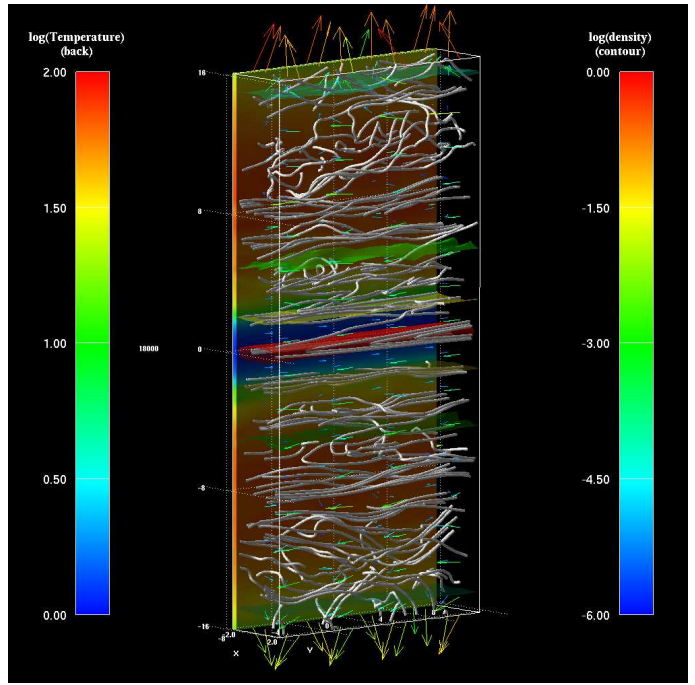}
\end{center}
\caption{Snapshot of the simulation with $\gamma=1.4$ at 180 rotation time 
($t\Omega/2\pi=180$). White lines indicate magnetic field lines, colors 
on the back show normalized temperature, $T/T_0$, ({\it left legend}), 
color contours denote iso-density surfaces ({\it right legend}), 
and arrows indicate velocity fields. }
\label{fig:snp}
\end{figure}

Continued from the previous section we examine the time evolutions 
of the box averaged $\langle \alpha \rangle_{x,y,z}$ in Figure \ref{fig:totalp}.
After the quasi-steady saturated states are achieved, 
the three cases show different behavior. The isothermal case ($\gamma=1.0$) 
show large fluctuations of $\langle \alpha \rangle_{x,y,z}$ from 0.015 to 0.04. 
On the other hand, the case with $\gamma=1.4$ exhibits much milder 
behavior with $\langle \alpha \rangle_{x,y,z}$ kept in a range of 0.01-0.015. 
The case with 
$\gamma=1.03$ shows intermediate behavior; although the increase of the
$\alpha$ is faster with showing large fluctuations in the earlier time, 
$2\pi t/\Omega\lesssim 120$, it settles down to a softer state later, 
between those for $\gamma=1.4$ and 1.0. 
Interestingly enough, the similar trends are observed in the disk winds, 
which we discuss the details in \S \ref{sec:vflow}.

Figure \ref{fig:snp} exhibits a snapshot structure of the case with 
$\gamma=1.4$ at 180 rotation time ($t\Omega/2\pi=180$). 
The turbulent magnetic field, mostly dominated by the toroidal component, 
is amplified by MRI and winding due to the differential rotation. 
The temperature contour on the back shows that the temperatures in 
$4\lesssim|z/H_0|\lesssim 12$ 
increase up to more than 50 times of the initial value, while the 
temperature in the midplane does not increase so much. 
$\frown$-shaped field lines, which are typical for Parker 
(magnetic buoyancy) instability \citep{par55,nis06} are seen in the surface 
regions, and the vertical outflows 
are observed from both the upper and lower surfaces, similarly to those seen 
in the isothermal simulations \citep{suz09,suz10}.

\subsection{Coronae}
\begin{figure}[h]
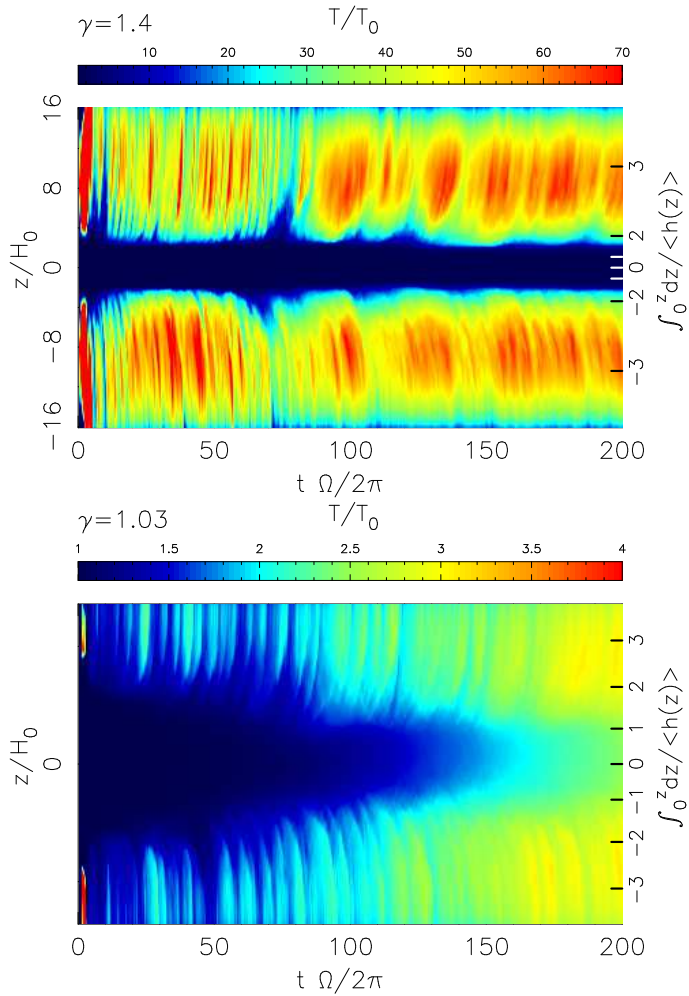

\begin{center}
\includegraphics[width=0.5\textwidth]{t-z_gamma14_T1.ps}\\
\includegraphics[width=0.5\textwidth]{t-z_gamma103_T1.ps}
\end{center}
\caption{Time-distance diagrams of temperatures, $\langle T/T_0\rangle_{x,y}$,
of the cases with $\gamma=1.4$ ({\it upper}) and 1.03 ({\it lower}). 
The labels for the horizontal and vertical axes are the same as in 
Figure \ref{fig:tzalpha}. 
}
\label{fig:tzT}
\end{figure}

\begin{figure}[h]
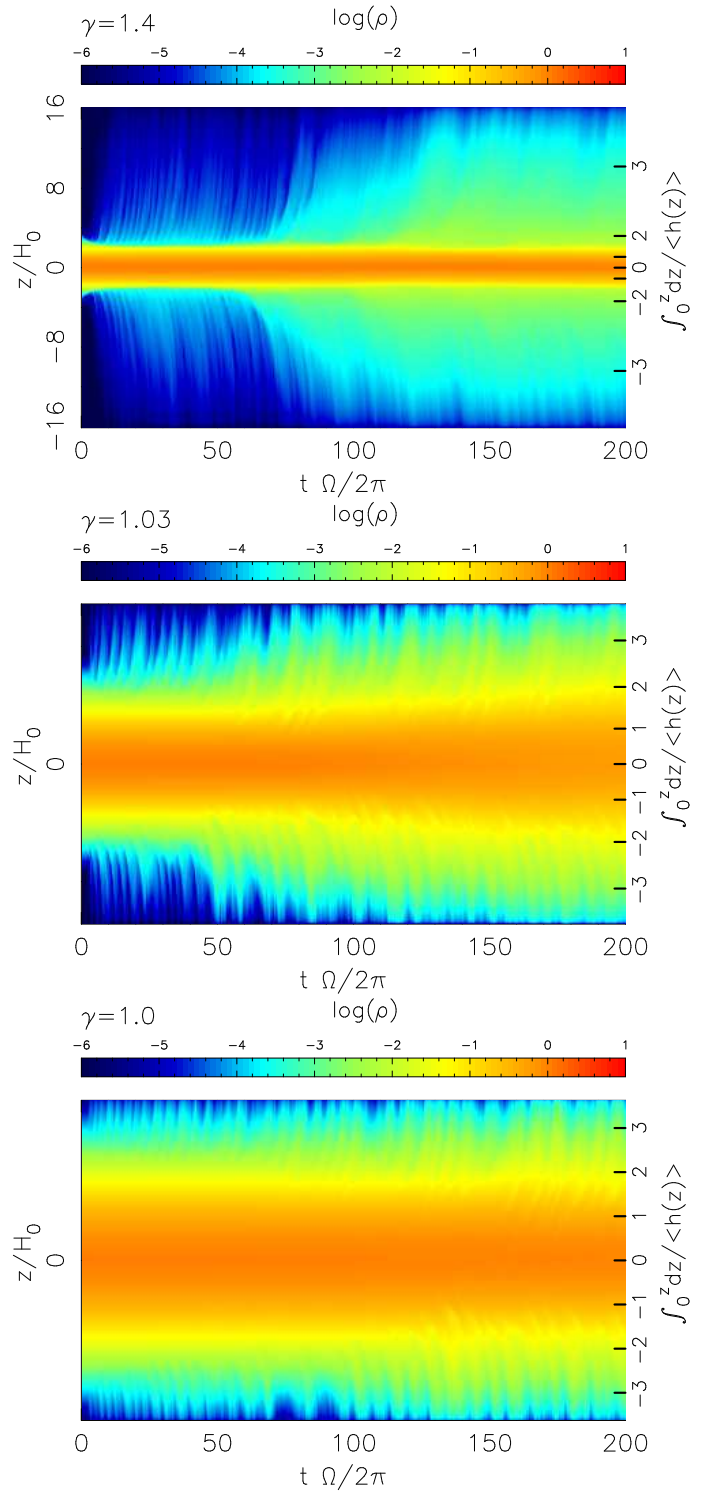

\begin{center}
\includegraphics[width=0.5\textwidth]{t-z_gamma14_rh1.ps}\\
\includegraphics[width=0.5\textwidth]{t-z_gamma103_rh1.ps}\\
\includegraphics[width=0.5\textwidth]{t-z_gamma10_rh1.ps}
\end{center}
\caption{Time-distance diagrams of densities, $\langle\rho\rangle_{x,y}$,
of the cases with $\gamma=1.4$ ({\it top}), 1.03 ({\it middle}),
and 1.0 ({\it bottom}). 
The labels for the horizontal and vertical axes are the same as in 
Figure \ref{fig:tzalpha}. 
} 
\label{fig:tzrho}
\end{figure}

Figures \ref{fig:tzT} and \ref{fig:tzrho} present $t$--$z$ diagrams of 
the temperatures and the densities of the three cases with $\gamma =1.4, 1.03$, 
\& 1.0, whereas we do not show the temperature of the isothermal 
($\gamma=1.0$) case. As shown in Figure $\ref{fig:tzT}$ as well as 
Figure \ref{fig:snp}, the gas in the surface regions of the non-isothermal 
cases is heated up. The heating is done by the dissipation of the magnetic 
energy which is amplified by the MRI and the winding due to the differential 
rotation. Since the simulations do not explicitly take into account 
resistivity, the dissipation of the magnetic field occurs numerically in 
the sub-grid scales, whereas we expect that this actually takes place by 
cascading of small-scale turbulence. 

The two cases of Figure \ref{fig:tzT} show an initial temperature rise 
in $t\Omega/2\pi<5$, 
because the initial densities in the upper regions, $|z/H_0|>4.55$, are larger 
than the hydrostatic value (Equation \ref{eq:initrho}); the gas initially 
flows down from there toward the midplane, which causes the initial heating. 
The entire region is eventually relaxing to the quasi-steady state, 
which we analyze from now. In both the cases, the regions of $2.5<
|\int_0^z\frac{dz}{\langle h(z)\rangle}_{\Delta t_{\rm ave},x,y}|<3$ are most 
effectively heated up to $T/T_0\sim 60$ in the case with $\gamma=1.4$ and 
$T/T_0\sim 3$ in the case with $\gamma=1.03$. 
Because the heating is done by the dissipation of the magnetic field which is 
amplified by the MRI and the winding, the increase of the temperature follows 
the increase of $\langle \alpha \rangle_{x,y}$ shown in Figure 
\ref{fig:tzalpha}. This is more easily seen in the case with $\gamma=1.03$; 
the temperature increases in $t\Omega/2\pi \gtrsim 150$, which is delayed 
compared to the increase of $\langle \alpha \rangle_{x,y}$. 

The top panel of Figure \ref{fig:tzT} illustrates that the hot regions with 
$T/T_0>50$ are clearly separated from the cool midplane region with $T/T_0 
\ll 10$ by the transition regions located at $\int_0^z\frac{dz}
{\langle h(z)\rangle_{\Delta t_{\rm ave},x,y}}\approx \pm 1.5$; from now we call these
hot regions coronae. On the other hand, in the small $\gamma$
($=$1.03) case, the temperatures of the surface regions are not so 
high with $T/T_0\approx 3$ in the upper regions. 

\begin{figure*}[th]
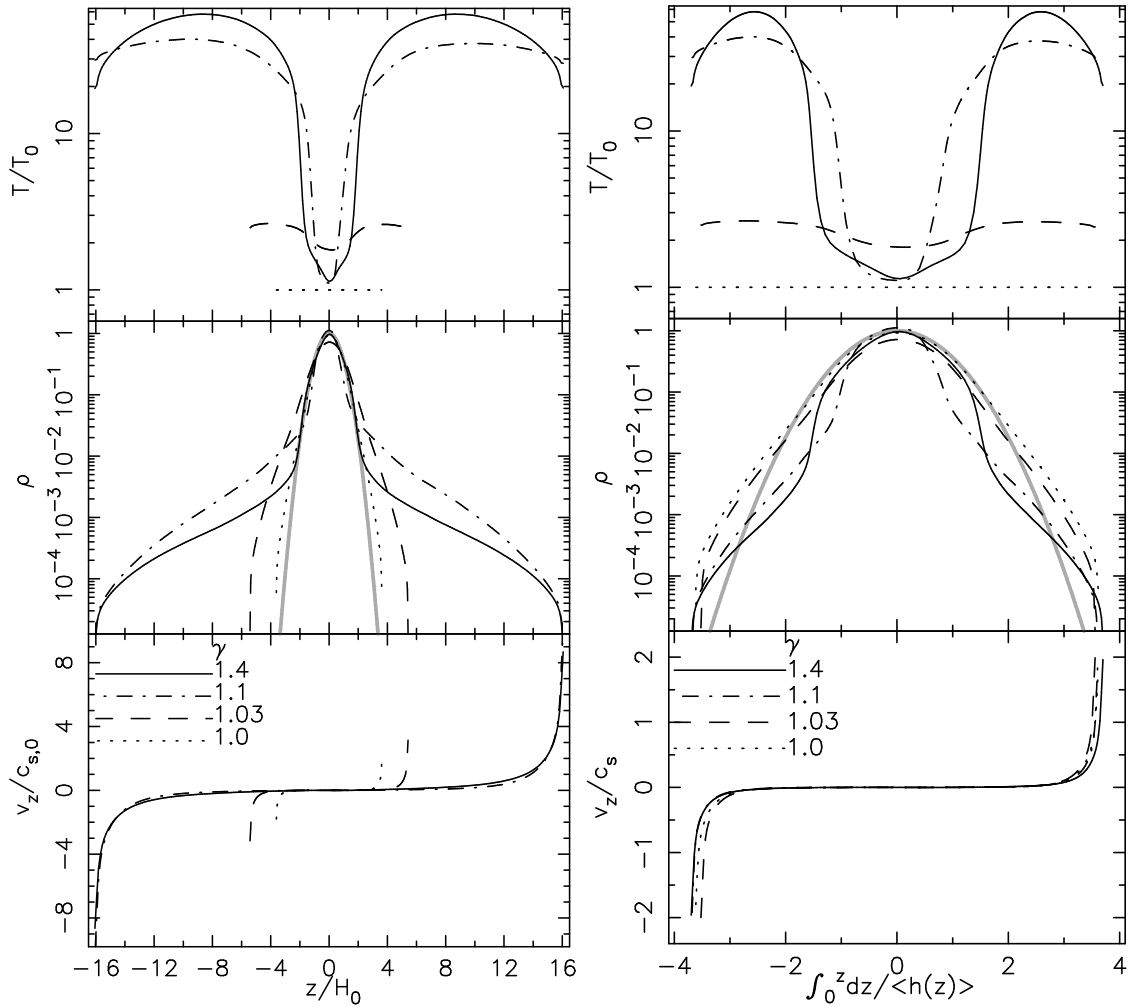

\begin{center}
\includegraphics[width=0.42\textwidth]{verstr_tav_wgamma_z03.ps}
\includegraphics[width=0.4\textwidth]{verstr_tav_wgamma_zH3.ps}
\end{center}
\caption{Comparison of the time and horizontally averaged vertical structures 
of the cases with $\gamma=1.4$ ({\it solid}), 1.1 ({\it dot-dashed}),
1.03 ({\it dashed}), and 1.0 ({\it dotted}). The horizontal axis is vertical 
distance in units of the initial scale height $z/H_0$ for the left panels 
and in units of the final scale heights, 
$\int_{0}^{z}dz/\langle h(z)\rangle_{\Delta t_{\rm ave},x,y}$, for the right panels.
From top to bottom, 
the temperatures normalized by the initial value, the densities, and the 
vertical velocities normalized by the initial constant sound speed, 
$c_{\rm s,0}$, ({\it left panels}) or normalized by the final local sound 
speeds, $c_{\rm s}(z)$, ({\it right panels}) are compared. 
The gray thick lines in the middle panels indicate the initial density 
profile. }
\label{fig:vst_zH0}
\end{figure*}

For more quantitative inspection we display the time and horizontally averaged 
vertical structures of the four cases, $\gamma=1.0,1.03,1.1$, \& 1.4 in 
Figure \ref{fig:vst_zH0}, where the time averages are taken over 
$\Delta t_{\rm ave}$ in Table \ref{tab:bsz} after the quasi-steady 
saturation is achieved (Figure \ref{fig:totalp}).
The top panel illustrates that the coronae with $T/T_0\approx 40$ also form 
in the case with $\gamma=1.1$.
We have found that the formation of the coronae with $T/T_0\approx 40-70$ 
is universal in the cases with $\gamma\ge 1.1$. In these cases, with the 
large increases of the temperature (around $\int_{0}^{z}dz/\langle h(z)
\rangle_{\Delta t_{\rm ave},x,y}\approx \pm 1.5$ for $\gamma=1.4$ and $\pm 1$ for 
$\gamma=1.1$) the density (middle panels) rapidly decreases to keep 
the nearly pressure-balanced structure. We call this pressure balanced 
structure with temperature rise a transition region. 

The existence of the hot coronae above the cool midplane region 
is quite different from some results of recent stratified 
shearing box simulations.  For instance, 
\citet{bod12} performed the simulations with fixing the temperatures to 
the initial value ($T=T_0$) and the velocities to zero ($v_z=0$) 
at the top and bottom $z$ boundaries. They showed that the maximum temperature 
is obtained at the midplane and the temperature monotonically decreases 
to the surfaces. This shows that the boundary condition at the $z$ 
boundaries play a significant role in the vertical temperature and velocity 
structures. 

Coupled with the formation of the hot coronae ($\gamma\ge 1.1$) or the warm  
regions ($1<\gamma<1.1$), the gas is lifted up from the midplane region 
to the upper regions as shown in Figure \ref{fig:tzrho}. 
In the case with $\gamma=1.4$ (top panel), it is 
clearly seen that the coronae are gradually filled with denser material 
in $t\Omega/2\pi\gtrsim 50$ by the evaporation from the midplane region, 
and at later times the evaporated gas is mainly supported by the gas pressure 
of the hot coronae (\S \ref{sec:vflow}).  
The time averaged density structures of the cases 
$\gamma=1.1$ and 1.4 (middle panels of Figure \ref{fig:vst_zH0}) 
quantitatively show the large uplifted gas compared with the initial 
hydrostatic profile with the constant $T_0$ (gray thick lines). 
The supply of the gas to the coronal regions is also seen in the case with 
$\gamma=1.03$ (middle panel of Figure \ref{fig:tzrho}) although the amount 
is smaller than those of larger $\gamma$ cases. 
Even in the isothermal case (bottom panel of Figure \ref{fig:tzrho}) 
quasi-periodic uplifting motions of the gas are observed, which are by the 
Poynting flux associated with the breakups of channel flows \citep{suz09}. 
The uplifted gas is finally connected to disk winds as will be discussed 
in \S \ref{sec:vflow}.



\subsection{Saturation of Magnetic Field}
\label{sec:satB}

\begin{figure*}[th]
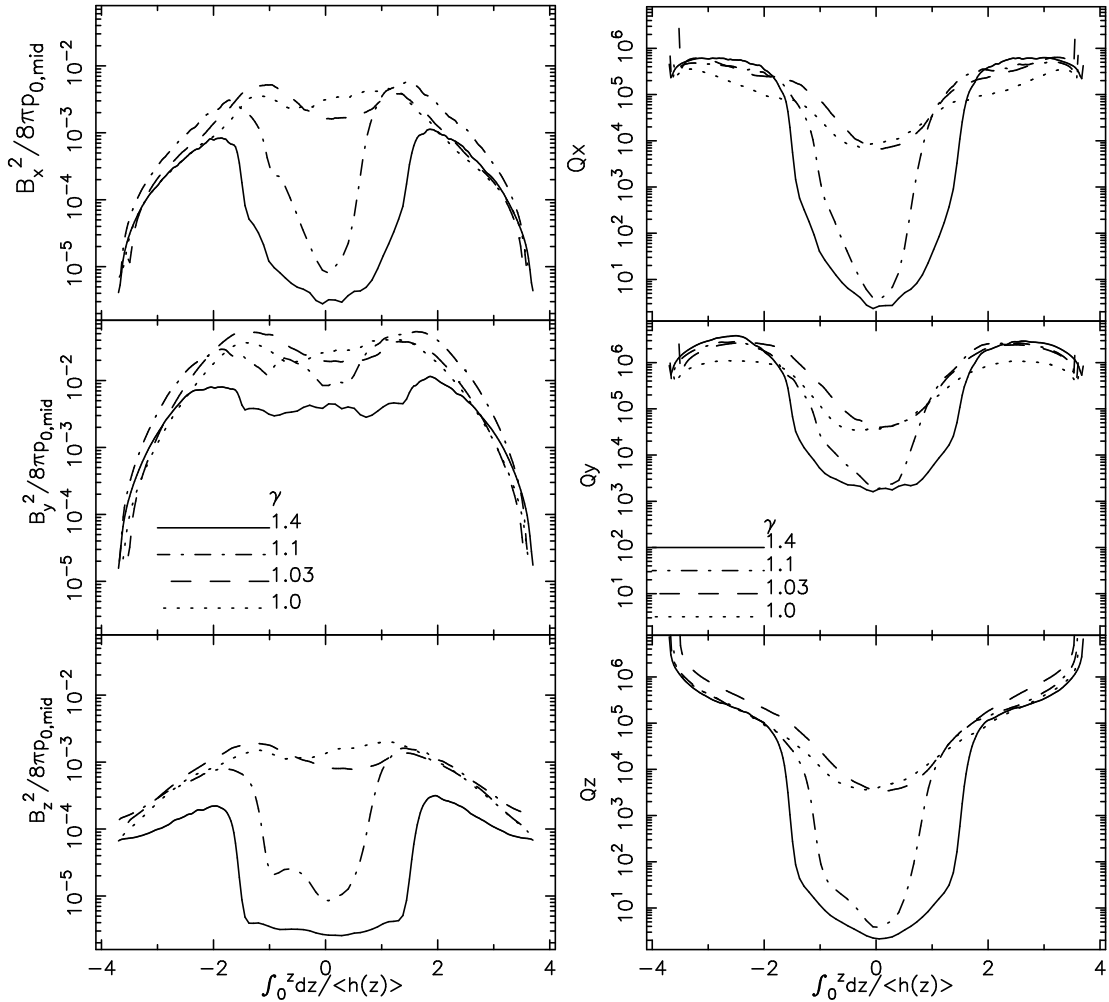

\begin{center}
\includegraphics[width=0.407\textwidth]{verstr_tav_wgamma_zH_B1.ps}
\includegraphics[width=0.4\textwidth]{verstr_tav_wgamma_zH_Q2.ps}
\end{center}
\caption{Comparison of the time and horizontally averaged vertical structures 
of the $i=x$ ({\it top}), $y$ ({\it middle}), and $z$ ({\it bottom}) components 
of magnetic energies, $B_i^2/8\pi$, normalized by the initial gas pressure at 
the midplane, $p_{\rm 0,mid}$ ({\it left}), and quality factors, $Q_i$, 
for MRI ({\it right}) of the cases with $\gamma=1.4$ 
({\it solid}), 1.1 ({\it dot-dashed}), 1.03 ({\it dashed}), and 1.0 
({\it dotted}). The horizontal axis, 
$\int_{0}^{z}dz/\langle h(z)\rangle_{\Delta t_{\rm ave},x,y}$, is measured 
in the final scale heights.}
\label{fig:vst_zHt_B}
\end{figure*}

\begin{figure}[h]
\begin{center}
\includegraphics[width=0.4\textwidth]{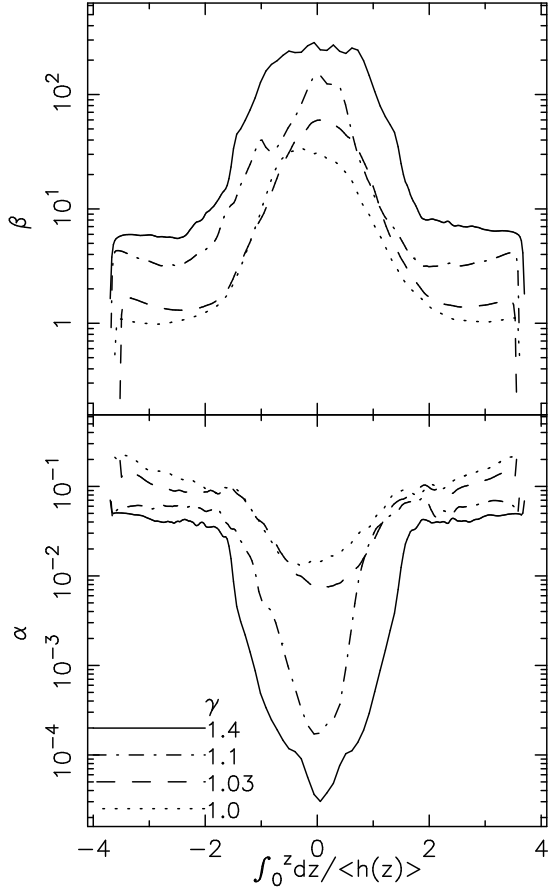}
\end{center}
\caption{Comparison of the time and horizontally averaged vertical structures 
of the plasma $\beta$ values ({\it upper}) and the $\alpha$ values 
({\it bottom}) of the cases with $\gamma=1.4$ ({\it solid}), 1.1 
({\it dot-dashed}), 1.03 ({\it dashed}), and 1.0 ({\it dotted}). 
The horizontal axis, $\int_{0}^{z}dz/\langle h(z)\rangle_{\Delta t_{\rm ave},x,y}$, 
is measured in the final scale heights. 
}
\label{fig:vst_zHt_ab}
\end{figure}

Magnetic field plays a central role in the formation of the coronae because 
its dissipation leads to the heating. The left panels of 
Figure \ref{fig:vst_zHt_B} compare the time and horizontally averaged magnetic 
energies of the four cases with $\gamma=1.4$, 1.1, 1.03, and 1.0.  
The toroidal ($y$) component dominates as expected because of the winding 
by the differential rotation. 
In the surface regions, $|\int_{0}^{z}dz/\langle h(z)\rangle_{\Delta t_{\rm ave},x,y}|
\gtrsim 2$, these four cases show similar profiles one another. 
On the other hand, in the midplane region larger $\gamma$ cases give lower 
magnetic energies of the $x$ and $z$ components in particular. $B_y^2$ only 
weakly depends on $\gamma$ because the strength of $B_y$ is mostly controlled 
by the winding. As a result, in the midplane region of larger $\gamma$ cases 
more or less coherent toroidal magnetic field dominates (Figure \ref{fig:snp}). 

The lower levels of $B_x^2$ and $B_z^2$ are mainly because of the 
insufficient resolution. 
The right panels of Figure \ref{fig:vst_zHt_B} display the time and 
horizontally averaged 
$i$-th components ($i=x,y,z$) of quality factors with respect to MRI 
\citep{nob10,haw11}, 
\begin{equation}
Q_i = 2\pi \frac{\sqrt{v_{{\rm A},i}^2}}{\Omega \Delta l_i} 
= 2\pi\sqrt{\frac{B^2_{i}}{4\pi\rho}} \frac{1}{\Omega \Delta l_i},
\end{equation} 
where $v_{{\rm A},i}=B_i/\sqrt{4\pi \rho}$ is \Alfven velocity along with an 
$i$--th direction and $\Delta l_i(=\Delta x,\Delta y \Delta z)$ is the grid 
size of an $i$--th component. 
$Q_i$ measures the number of mesh points which resolves 
the most unstable wavelength of MRI. According to \citet{san04}, 
$Q_z\gtrsim 6$ is a necessary condition for a vertical magnetic field to 
get a linear growth rate close to the analytic prediction from MRI. 
$Q_x$ and $Q_z$ of the cases with $\gamma=1.4$ and 1.1 at the midplane is 
small $<5$. In these cases the initial scale height, $H_0$, is resolved by 
8 mesh points.
Although at later time one scale height, $h(z)$, is eventually 
resolved by a larger number of grids, $\sim 10-15$, it is still insufficient 
to capture fine turbulent structure by the MRI at the midplane 
\citep{sim09,haw11,haw13,pb13a,pb13b}. 
Therefore, the magnetic field strength at the midplane 
in the large $\gamma\ge 1.1$ is supposed to be underestimated and not to 
develop to the physically saturated state. 
On the other hand, in the small 
$\gamma< 1.1$ cases and the surface regions of the large $\gamma$ cases 
we can safely discuss the saturation of the magnetic fields. 

Figure \ref{fig:vst_zHt_ab} displays the time and horizontally averaged 
plasma $\beta$ values, 
$$\langle \beta \rangle_{\Delta t_{\rm ave},x,y} 
= \frac{8\pi\langle p \rangle_{\Delta t_{\rm ave},x,y}}
{\langle B^2 \rangle_{\Delta t_{\rm ave},x,y}}$$ (upper panel) 
and $\alpha$ values, 
$$\langle \alpha \rangle_{\Delta t_{\rm ave},x,y} 
= \frac{\langle \rho v_x \delta v_y - B_x B_y/4\pi
\rangle_{\Delta t_{\rm ave},x,y}}{\langle p\rangle_{\Delta t_{\rm ave},x,y}}$$ 
(lower panel). 
$\langle \beta\rangle_{t_{\rm ave},x,y}$ is larger and $\langle \alpha
\rangle_{t_{\rm ave},x,y}$ is smaller for smaller $\gamma$.  
In the surface regions, this is explained by the larger gas pressure due to 
the hotter coronae for larger $\gamma$ since the magnetic field strengths are 
similar among different $\gamma$ cases (left panels of 
Figure \ref{fig:vst_zHt_B}).  
The large $\beta$ and small $\alpha$ values at the midplane region in 
the cases with $\gamma=1.4$ and 1.1 are mainly because of the weaker magnetic 
field strength owing to the insufficient resolution 
(Figure \ref{fig:vst_zHt_B}). 

In the surface regions, 
$|\int_{0}^{z}dz/\langle h(z)\rangle_{\Delta t_{\rm ave},x,y}|\gtrsim 2$, the $\langle 
\beta\rangle_{\Delta t_{\rm ave},x,y}$ values of the different cases are almost 
spatially constant (Figure \ref{fig:vst_zHt_ab}), which indicates 
that the magnetic energy decreases with increasing $|z|$ (left panels of 
Figure \ref{fig:vst_zHt_B}) in the same manner as the decrease of the gas 
pressure, whereas the 
level of $\langle \beta\rangle_{\Delta t_{\rm ave},x,y}$ depends on $\gamma$ 
reflecting the coronal gas pressure in the numerator of $\beta$. 
In other words, the magnetic energy dissipates in these surface regions; 
the magnetic energy becomes relatively 
important with an elevating altitude from the midplane with decreasing 
density, and at the locations, $\int_{0}^{z}dz/\langle h(z)
\rangle_{\Delta t_{\rm ave},x,y}\approx \pm 2$, with $\rho \approx$ 
$10^{-2}$--$10^{-3}$ times $\rho_{\rm mid}$ (Figure \ref{fig:vst_zH0}) 
and $1\lesssim\beta< 10$, {\it i.e.}, nearly equipartition, 
the magnetic energy starts to be gradually converted to other forms. 
This is the source of the coronal heating and the driving of the disk winds.  
Although in our simulations the dissipation of the magnetic energy is due to
numerical dissipation at the sub-grid scales, in reality the MHD turbulence 
is supposed to dissipate through the energy cascading to smaller scales.

\subsection{Vertical Outflows}
\label{sec:vflow}

\begin{figure*}[t]
\begin{center}
\includegraphics[width=0.8\textwidth]{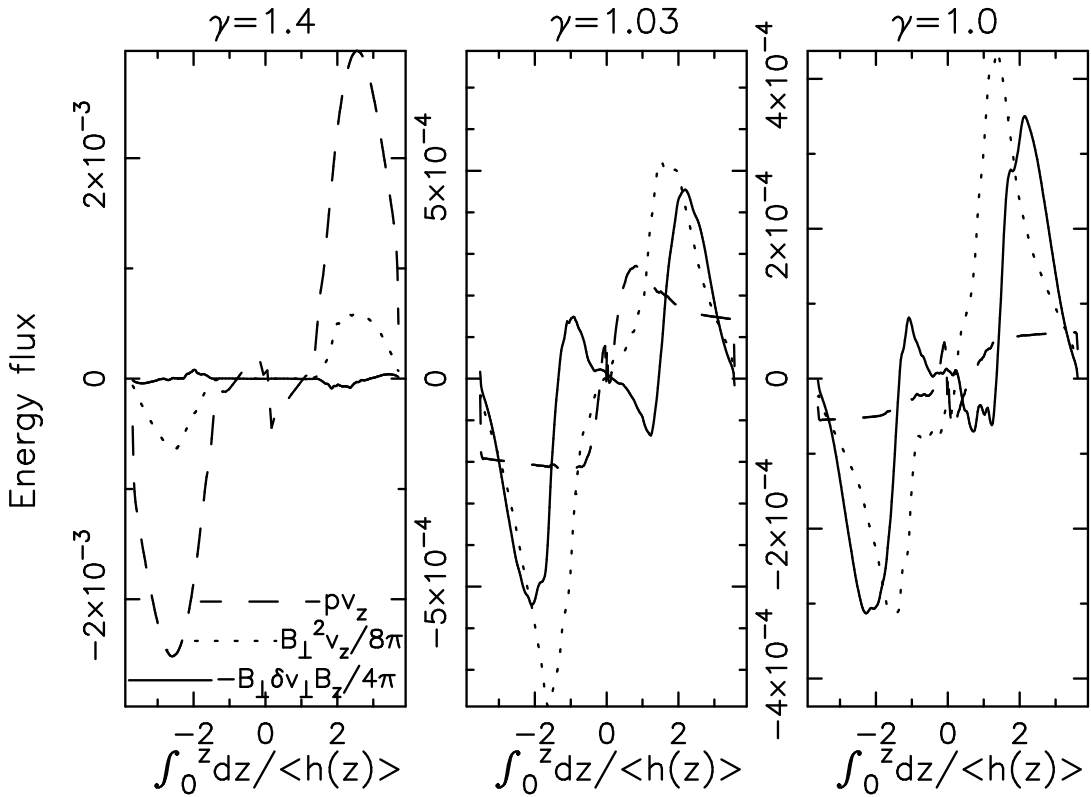}
\end{center}
\caption{Comparison of the time and horizontally averaged vertical structures 
of the gas pressure ({\it dashed}), the magnetic pressure ({\it dotted}), and 
the magnetic tension ({\it solid}) terms of the cases with $\gamma=1.4$ 
({\it left}), 1.03 ({\it middle}), and 1.0 ({\it right}). See {\it text} 
(Equations \ref{eq:engEu} \& \ref{eq:vefx} and their explanations) for details.}
\label{fig:efx}
\end{figure*}

The bottom panels of Figure \ref{fig:vst_zH0} show that vertical outflows 
are accelerated in the surface regions. The left bottom panel 
($v_z/c_{\rm s,0}$--$z/H_0$) shows that in larger $\gamma$ cases faster 
outflows are driven at higher altitudes. This is because the material is 
more extended to higher altitudes and the winds are  
effectively accelerated by the gas pressure of the hotter gas.  
If we plot the same quantity in the 
$v_z/c_{\rm s}(z)$--$\int_{0}^{z}dz/\langle h(z)\rangle_{\Delta t_{\rm ave},x,y}$ 
diagram (right bottom panel), the profiles of the shown four models resemble 
one another; the vertical velocities reach the local sound speeds at 
$\int_{0}^{z}dz/\langle h(z)\rangle_{\Delta t_{\rm ave},x,y}\approx \pm 3.5$ 
where the densities decrease to $\sim 10^{-4}-10^{-5}$ times $\rho_{\rm mid}$. 
By these outflows, the total mass in the simulation box of each case 
gradually decreases with time. Typically 5-10 \% of 
the initial mass is lost until the end of the simulations at $t\Omega/\pi
=200$ rotations (Table \ref{tab:bsz}). 


We inspect how the driving mechanism of these disk winds is different 
for different $\gamma$. 
By rearranging Equation (\ref{eq:eng}), 
we can derive an Eulerian form of the energy equation: 
\begin{eqnarray}
&&\frac{\partial}{\partial t}\left[\rho\frac{v^2}{2}+\rho e 
+ \frac{B^2}{8\pi} + \rho \Psi\right] \nonumber \\
&=&- \mbf{\nabla \cdot}
\left[\rho\mbf{v}\left(\frac{v^2}{2}+h +\Psi\right) -\frac{1}{4\pi}
(\mbf{v\times B})\mbf{\times B}\right] \nonumber \\
&=& - \mbf{\nabla \cdot}
\left[\left\{\mbf{v}\left(\rho\frac{v^2}{2}+\rho e + \frac{B^2}{8\pi} 
+\rho\Psi\right) \right\}\right. \nonumber \\
& &\left. + \left\{\left(p + \frac{B^2}{8\pi}\right)\mbf{v} 
- \frac{\mbf{B}}{4\pi}(\mbf{B\cdot v})\right\}\right]  ,
\label{eq:engEu}
\end{eqnarray}
where $\Psi \equiv \frac{\Omega^2}{2}(z^2-3x^3)$ and $h = e +p/\rho$. 
One can see from the last equality, the energy flux can be divided into 
the two parts;  the first $\{\;\}$ bracket indicates the energy advected with 
$\mbf{v}$ and the second $\{\;\}$ bracket indicates the work acting on gas; 
the second part exactly appears in the spatial derivative term in the 
Lagrangian form  
(Equation \ref{eq:eng}). Here, we examine this work component in the 
simulations. The $z$ component of the second part can be explicitly written as 
$$
\hspace{-1.5cm}
\left(p + \frac{B^2}{8\pi}\right)\mbf{v} - \frac{\mbf{B}}{4\pi}(\mbf{B\cdot v})
$$
\begin{equation}
= p v_z+ \frac{B_{\perp}^2}{8\pi}v_z - \frac{B_z}{4\pi}B_{\perp}v_{\perp} 
- \frac{B_z^2}{8\pi}v_z, 
\label{eq:vefx}
\end{equation}
where $B_{\perp}^2=B_x^2+B_y^2$ and $B_{\perp}v_{\perp} = B_x v_x + B_y v_y$.
The first, second, and third terms indicate the work by gas pressure, 
magnetic pressure, and magnetic tension, respectively. The last term is 
canceled out by a term from the advection component. We compare the first --
third terms of the three cases with $\gamma=1.4$ (left panel), 1.03 (middle 
panel), and 1.0 (right panel) in Figure \ref{fig:efx}.

If a line in Figure \ref{fig:efx} decreases with increasing $z$ in the $z>0$ 
region, the force acts on gas to drive a vertical upflow, and vice versa in 
the $z<0$ region.
In the isothermal ($\gamma=1.0$) case, the magnetic pressure (dotted line) 
and tension (solid line) comparably contribute to driving the vertical 
outflows, while the contribution from the gas pressure is quite small. 
The ``injection regions'' of the magnetic tension form around $\int_{0}^{z}
dz/\langle h(z)\rangle_{\Delta t_{\rm ave},x,y} \approx \pm 1.5$ as pointed out by \citet{suz09}; 
from these regions, the Poynting fluxes associated with the tension are 
injecting toward both the midplane and surface directions.  

On the other hand, in the case with $\gamma=1.4$, the gas pressure largely 
dominates the magnetic components in driving the disk winds. In this case, 
the magnetic energy in low-altitude regions once dissipates to heat up 
the gas. The gas pressure, which increases owing to the magnetic heating, 
finally contributes to driving the vertical outflows. This is in contrast 
to the isothermal case, in which the magnetic forces directly drive the 
vertical outflows. 
The behavior of the case with $\gamma=1.03$ lies between the two cases. 
While the largest contribution is from the magnetic pressure, the magnetic 
tension and the gas pressure also play a significant role.

Inspecting all the cases with different $\gamma$, we can conclude the following 
results on the driving mechanisms of the disk winds: 
While in the isothermal and small 
$\gamma\lesssim 1.03$ cases the vertical outflows are mainly driven by the 
Poynting flux, in the large $\gamma\gtrsim 1.1$ cases the gas pressure 
dominantly drives the vertical outflows. However, we should note that, 
even in the large $\gamma$ cases the magnetic fields play an important 
role, because the gas pressure is maintained by the 
dissipation of the magnetic energy which is amplified by the MRI and the 
winding due to the differential rotation. 
Since the disk winds are driven from the surface regions where the MRI is 
well-captured, increasing the 
numerical resolution does not change the obtained properties of the disk 
winds so much, 
although the magnetic field at the midplane of the large $\gamma$ cases 
would be affected because the numerical resolution there is insufficient 
at the moment (Figure \ref{fig:vst_zHt_B}).  

Magnetic buoyancy \citep{par55} is also considered to operate in the 
surface regions and contribute to the outflows \citep{nis06,mac13}. 
As discussed in Figure \ref{fig:snp}, 
$\frown$--shaped field lines, typical for the Parker instability, are 
sometimes observed. Compared to the isothermal \citep[see also][]{suz10} 
and small $\gamma$ cases, however, the contribution from the magnetic buoyancy 
is smaller in the large $\gamma$ cases. In general, the Parker instability 
sets in for magnetically dominated (small $\beta$) condition 
with strong stratification (short pressure scale height). In the large 
$\gamma\ge 1.1$ cases the time and horizontally averaged $\beta$ is not 
so small $\approx 5$ (Figure \ref{fig:vst_zHt_ab}) and the scale height 
in the surface regions is not short, which tend to suppress the Parker 
instability compared to the small $\gamma$ cases.

\begin{figure}[h]
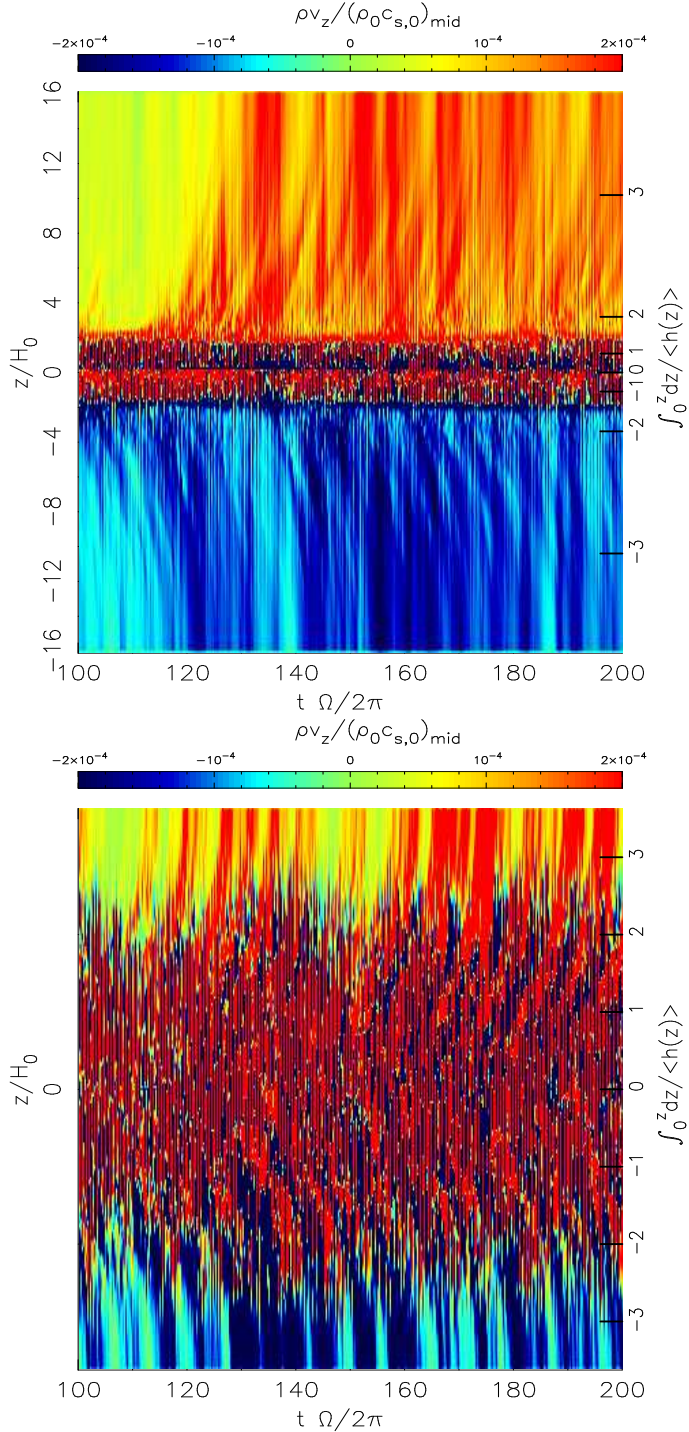

\begin{center}
\includegraphics[width=0.5\textwidth,height=0.4\textheight]{t-z_gamma14_rhovz1.ps}\\
\includegraphics[width=0.5\textwidth,height=0.4\textheight]{t-z_gamma10_rhovz1.ps}
\end{center}
\caption{Time-distance diagrams of the normalized mass flux, $\rho v_z/(\rho_0 
c_{\rm s,0})_{\rm mid}$ 
of the cases with $\gamma=1.4$ ({\it upper}) 
and 1.0 ({\it lower}). 
The labels for the horizontal and vertical axes are the same as in 
Figure \ref{fig:tzalpha}. 
}
\label{fig:tzrhovz}
\end{figure}

The difference of the driving mechanisms influences the time-dependency of the 
disk winds. Figure \ref{fig:tzrhovz} compares the $t-z$ diagrams of 
the mass flux of the cases with $\gamma=1.4$ (upper panel) and 1.0 (lower 
panel). The two cases show quite different appearances. The isothermal 
case (lower panel) shows a clearer on-off nature of the disk winds. 
As reported in \citet{suz09}, it is related to the quasi-periodic breakups
of channel-mode flows with 5-10 rotation times. 
In the case with $\gamma=1.4$, the mass flux increases to the saturated 
value in $t\Omega/2\pi \gtrsim 120$, after the gas is supplied to the coronal 
regions (top panel of Figure \ref{fig:tzrho}).
After that $\rho v_z$ exhibits smoother structure in the coronal regions, 
$|\int_0^z dz/\langle h(z)\rangle_{\Delta t_{\rm ave},x,y}|>1.5$, 
with rather quasi-steady vertical outflows. The midplane region 
($|\int_0^z dz/\langle h(z)\rangle_{\Delta t_{\rm ave},x,y}|<1.5$) 
with $T/T_0<10$ (Figure \ref{fig:vst_zH0}) seems to be separated from the 
coronal regions by the transition regions at 
$\int_0^z dz/\langle h(z)\rangle_{\Delta t_{\rm ave},x,y}\approx \pm 1.5$.
The quasi-steady nature of the 
disk winds in this case is related to the fact that the vertical outflows 
are mainly driven by the gas pressure. 
The coronal temperatures are more or less uniformly distributed in 
$|\int_0^z dz/\langle h(z)\rangle_{\Delta t_{\rm ave},x,y}|\gtrsim 2$, 
and the force by the gas pressure gradient are more time-steady,  
which is in contrast to the strong intermittency of the Poynting flux-driven 
outflows in the isothermal case. 

\subsection{Dependence on $\gamma$}

\begin{figure}[h]
\begin{center}
\includegraphics[height=0.43\textheight]{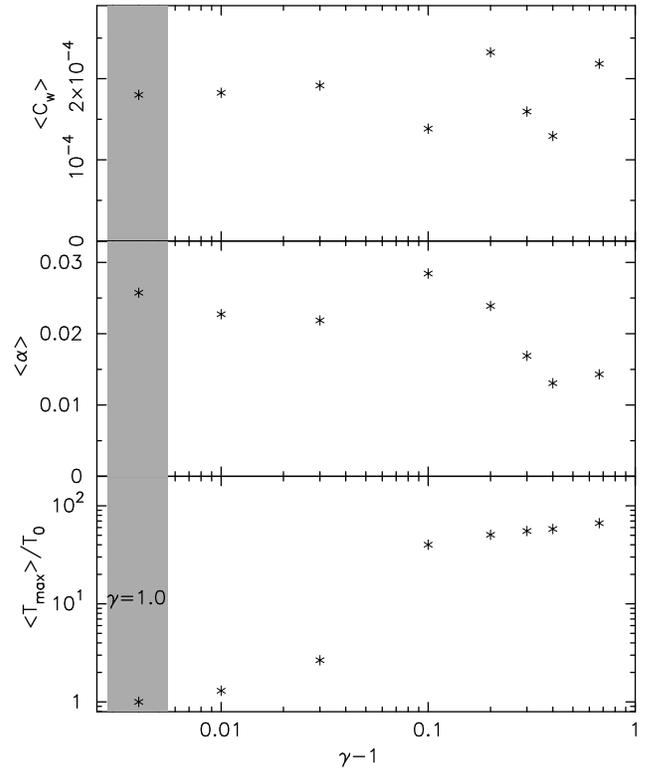}
\end{center}
\caption{Dependences of time averaged three quantities on 
$\gamma-1$. Shown are the sum of the nondimensionalized mass flux of the 
vertical outflows, $\langle C_{\rm w}\rangle_{\Delta t_{\rm ave},x,y}$, 
from the upper and lower surfaces (eq.\ref{eq:msflx}; {\it top}), 
the peak temperature, the sum of the box-averaged Maxwell and Reynolds 
stresses, $\langle \alpha \rangle_{\Delta t_{\rm ave},x,y,z}$, ({\it middle}), and 
$\langle T_{\rm max}\rangle_{\Delta t_{\rm ave},x,y}$ ({\it bottom}). 
In the shaded region at the left most location the results 
of the isothermal case ($\gamma=1.0$) are plotted.}
\label{fig:btave}
\end{figure}

We summarize typical time-averaged quantities of the simulations with 
different $\gamma$. The top panel of Figure \ref{fig:btave} shows the 
dependence of the mass flux of the disk winds. The shown quantity is 
\begin{equation}
\langle C_{\rm w}\rangle_{\Delta t_{\rm ave},x,y} = \left[
\langle(\rho v_z)_{\rm top}\rangle + \langle(-\rho v_z)_{\rm bot}\rangle\right] / 
\langle (\rho c_{\rm s})_{\rm mid}\rangle,
\label{eq:msflx}
\end{equation}
where the subscripts, `top' and `bot', indicates the top and bottom 
boundaries of the simulation box, and the subscript `mid' stands for 
the midplane. The variables in the brackets on the right hand side are also 
time and horizontally averaged. 
The meaning of Equation (\ref{eq:msflx}) would be clear, 
the sum of the mass fluxes from the upper and lower simulation boundaries, 
which is further normalized by the time and horizontally averaged 
$\rho c_{\rm s}$ at the midplane. 
Note that the $-$ sign for the mass flux from the bottom surface 
is to pick up the outflowing direction. 

The data points show that the nondimensional mass flux, $C_{\rm w}$, 
seems to be almost independent from $\gamma$ and is distributed within a 
factor of 2, which is consistent with the trend obtained from 
the time-averaged vertical structure (bottom right panel of Figure 
\ref{fig:vst_zH0})\footnote{Furthermore, since the normalization, 
$(\rho c_{\rm s})_{\rm mid}$, only weakly depends on $\gamma$ (see top and middle 
panels of Figure \ref{fig:vst_zH0}), the time-averaged mass flux itself 
is almost independent from $\gamma$}. 
This indicates that the {\it time-averaged} $C_{\rm w}$ dose not depend on 
the properties of the vertical 
outflows, either Poynting flux-driven winds with strong intermittency
(smaller $\gamma$) or more time-steady gas pressure-driven wind 
(larger $\gamma$), which is 
a little surprising. We suppose that the main reason of the insensitive 
$C_{\rm w}$ to $\gamma$ is 
that the original source of the vertical outflows is the magnetic energy. 
As discussed in \S \ref{sec:satB} the magnetic energy gradually dissipates 
to be converted to other forms of energies
at the locations, $\int_{0}^{z}dz/\langle h(z)
\rangle_{\Delta t_{\rm ave},x,y}\approx \pm 2$, which is almost independent from 
$\gamma$, with $\rho \approx$ $10^{-2}$--$10^{-3}$ times $\rho_{\rm mid}$ 
(Figure \ref{fig:vst_zH0}) and $1\lesssim\beta< 10$ (nearly equipartition; 
Figure \ref{fig:vst_zHt_ab}) .

In the small $\gamma$ regime the magnetic pressure and tension directly 
accelerate the Poynting flux-driven vertical outflows, while in the large 
$\gamma$ regime the magnetic energy 
is firstly transferred to the internal energy, the coronal heating in other 
words, and then the disk winds are driven by the gas pressure of the hot 
coronae. The locations (at $\rho \approx 10^{-2}-10^{-3}\rho_{\rm mid}$) of the 
energy conversion regulate the final mass flux, $C_{\rm w}\approx 2\times 
10^{-4}$, which is insensitive to $\gamma$.

The relative comparison of $C_{\rm W}$ among different $\gamma$ cases of 
the simulations is meaningful since the vertical box sizes are tuned to 
give 7.1--7.4 in units of the final scale heights (Table \ref{tab:bsz}). 
However, we should cautiously note that the absolute values 
of $C_{\rm W}$ should be taken with cares because they depend on the vertical 
box sizes \citep{suz10,fro13}; a larger vertical box would give a smaller 
$C_{\rm W}$.  
  
The middle panel of Figure \ref{fig:btave} presents the box- and time-averaged 
$\langle \alpha\rangle_{\Delta t_{\rm ave},x,y,z}$ values. 
We do not find any monotonic trend of 
$\langle \alpha\rangle_{\Delta t_{\rm ave},x,y,z}$
with $\gamma$ and the values are typically $(1-3)\times 10^{-2}$. 
However, we cannot proceed detailed saturation arguments 
\citep[e.g.][]{sim09,haw11,pb13a}, since the numerical resolution is not 
sufficient particularly in the midplane region of the large $\gamma$ cases 
(right panels of Figure \ref{fig:vst_zHt_B}). 

The bottom panel of Figure \ref{fig:btave} compares the maximum temperatures, 
$T_{\rm max}$, normalized by the initial value, $T_0$, of the different 
$\gamma$ cases. $T_{\rm max}$ is derived from the time and horizontally 
averaged vertical structure.
The derived $\langle T_{\rm max}\rangle_{\Delta t_{\rm ave},x,y}$ 
is located in a region of 
$2<|\int_{0}^{z}dz/\langle h(z)\rangle_{\Delta t_{\rm ave},x,y}|$ in the 
non-isothermal cases, 
whereas temperature becomes locally and transiently higher than $\langle 
T_{\rm max}\rangle_{\Delta t_{\rm ave},x,y}$. 
One can see a clear increasing trend with $\gamma$, because larger 
$\gamma$ simply 
corresponds to smaller net cooling (cooling - heating).
Moreover, $T_{\rm max}/T_0$ jumps up from $\gamma=1.01$ to 
$\gamma=1.1$, which corresponds to the change of the regime from the 
Poynting flux-driven winds to the gas pressure-driven winds.

\subsection{Wave Phenomena}

\begin{figure*}[t]
\begin{center}
\includegraphics[width=0.75\textwidth]{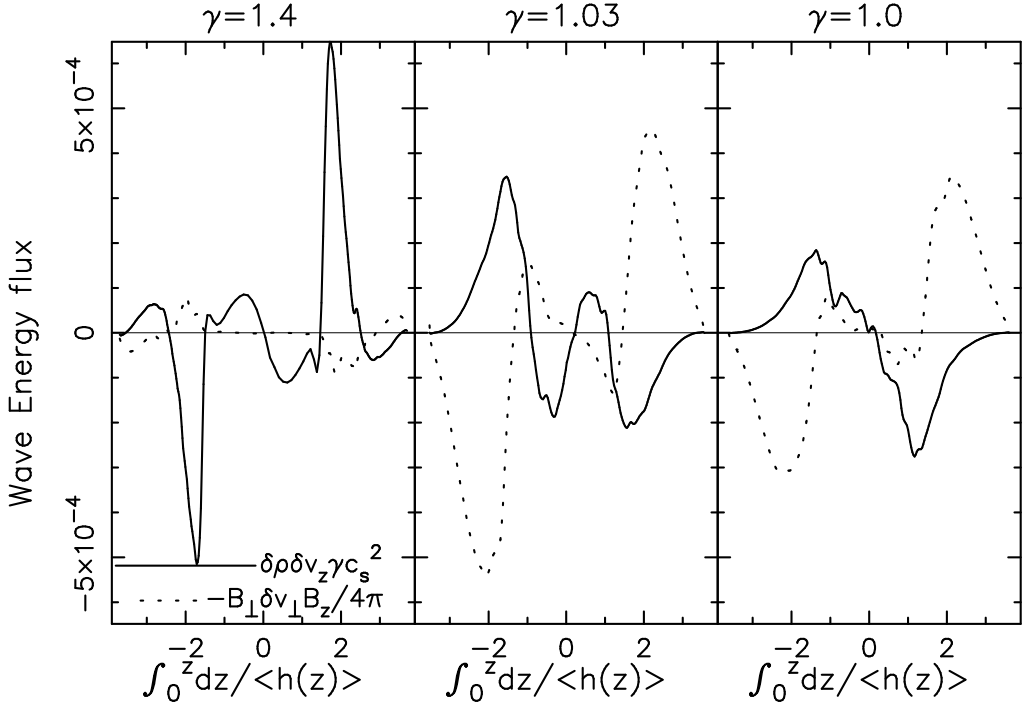}
\end{center}
\caption{Comparison of the time and horizontally averaged vertical structures 
of $-B_z\delta v_{\perp}B_{\perp}/4\pi$ ({\it dotted}) and 
$\delta\rho\delta v_{z}\gamma c_{s}^2$ ({\it solid}) for the cases with 
$\gamma=1.4$ ({\it left}), 1.03 ({\it middle}), and 1.0 ({\rm right}). 
They correspond to the net energy fluxes associated with 
\Alfvenic and sound-like waves, respectively. See {\it text} 
(Equations \ref{eq:Alf} \& \ref{eq:snd} and their explanations) for details.}
\label{fig:wave}
\end{figure*}

In \citet{suz09} we examined the vertical energy flux associated with 
wave-like activities and found \Alfvenic and sound-like waves propagating 
to both upward and downward directions. 
Following the procedure in \citet{suz09}, we inspect the 
vertical structures of two quantities, 
$-B_z\delta v_{\perp}B_{\perp}/4\pi$ and $\delta\rho\delta v_{z}\gamma c_{s}^2$. 
$-B_z\delta v_{\perp}B_{\perp}/4\pi$ is the Poynting flux of magnetic tension, 
as discussed in Equation (\ref{eq:engEu}), and can be rewritten as
\begin{equation}
-\frac{1}{4\pi}B_z\delta v_{\perp}B_{\perp}=\rho v_{{\rm A},z}(\delta 
v_{\perp,+}^2-\delta v_{\perp,-}^2),
\label{eq:Alf}
\end{equation}
where 
$\delta v_{\perp,\pm}=\frac{1}{2}(\delta v_{\perp}\mp 
B_{\perp}/\sqrt{4\pi\rho})$ are 
Els\"{a}sser variables, which correspond to the amplitudes of \Alfven waves 
propagating to the $\pm z$-directions.  
Thus, $-B_z\delta v_{\perp}B_{\perp}/4\pi$ corresponds to the net Poynting flux 
associated with propagating \Alfvenic disturbances to the $+z$ direction. 
$\delta\rho\delta v_z \gamma c_s^2$ is also rewritten as 
\begin{equation}
\delta\rho\delta v_z \gamma c_s^2=\rho \sqrt{\gamma}c_s
(\delta v_{\parallel,+}^2-\delta v_{\parallel,-}^2 ),
\label{eq:snd}
\end{equation}
where $\delta v_{\parallel,\pm}=\frac{1}{2}(\delta v_z\pm 
\sqrt{\gamma}c_s\frac{\delta\rho}{\rho})$ denote the amplitudes of sound 
waves\footnote{Strictly speaking, these are magnetosonic waves, 
namely the fast mode in the high $\beta$ plasma, and the slow 
mode that propagates along $z$ in the low $\beta$ plasma. 
Note also that the signs are opposite for $\delta v_{\perp,\pm}$ and 
$\delta v_{\parallel,\pm}$, reflecting the transverse and longitudinal 
characters.} propagating to the $\pm z$-directions. 
Here, note that the sound speed is expressed as $\sqrt{\gamma}c_{\rm s}$ 
since in this paper we define $c_{\rm s}$ as isothermal sound speed. 

Figure \ref{fig:wave} compares these quantities of the three cases 
with $\gamma=1.4$ (left panel), 1.03 (middle panel), and 1.0 (right panel).  
As already discussed in \citet{suz09}, in the isothermal case sound-like 
waves propagating to the midplane are observed. The peak values are 
located at the injection regions, $\int_{0}^{z}dz/\langle h(z)
\rangle_{\Delta t_{\rm ave},x,y}\approx \pm 1.5$ for 
the \Alfvenic waves (dotted line); \Alfvenic waves are injected from these 
regions mostly associated with the breakups of channel flows. 

The case with $\gamma=1.03$ exhibits similar structures except  
for $\delta\rho\delta v_z \gamma c_s^2$ (solid line) in the midplane region, 
$|\int_{0}^{z}dz/\langle h(z)\rangle_{\Delta t_{\rm ave},x,y}|<1$; the direction 
of the sound-like waves is upward (to both $\pm z$) from the midplane, 
which is opposite to that in the isothermal case. 
A speculative explanation is that the downward magnetic tension 
forces from the injection regions, which drive downward sound-like waves, 
are not relatively sufficient because the 
upward forces by the gas pressure are comparably important in this case 
(middle panel of Figure \ref{fig:efx}).

The case with $\gamma=1.4$ shows very different behavior. 
The energy flux of \Alfvenic waves is mostly dominated by that of  
sound-like waves.
In particular, it is nearly zero in the 
midplane region between the transition regions at
$\int_{0}^{z}dz/\langle h(z)\rangle_{\Delta t_{\rm ave},x,y} \approx \pm1.5$, 
which separate the cool midplane from the above hot coronae. 
The midplane region is protected 
from the magnetic perturbations in the upper coronae because the \Alfvenic 
disturbances are reflected at the transition regions (see \S\ref{sec:By}).    
The direction of the sound-like waves in this region is to the midplane. 
The large jumps are also seen in the sound-like waves at the transition 
regions; 
the sound speed also changes abruptly there owing to the change 
of the temperature, which causes the reflection of the sound-like waves.  
Thus, the sound-like waves are confined in the cool midplane region.

\begin{figure}[h]
\begin{center}
\includegraphics[width=0.4\textwidth]{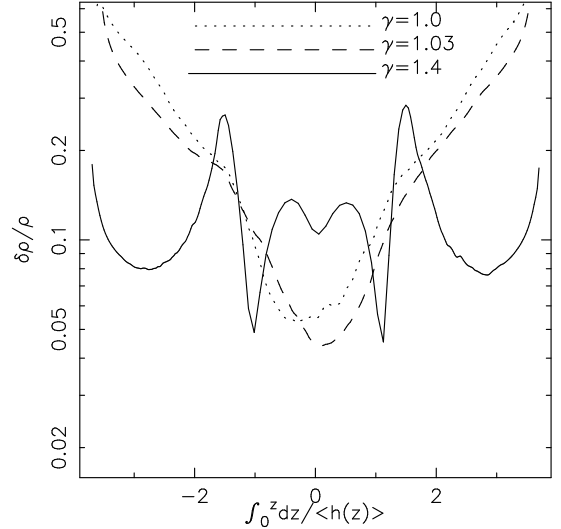}
\end{center}
\caption{Comparison of the time and horizontally averaged nondimensional 
density fluctuations, $\delta \rho/\rho$, of cases 
with $\gamma=1.4$ ({\it solid}), 1.03 ({\it dashed}), and 1.0 ({\it dotted}).}
\label{fig:ddns}
\end{figure}

Sound waves are associated with density perturbations. Therefore, we expect
that the different structures of $\delta\rho\delta v_z \gamma c_s^2$ 
will give different density perturbations. 
In Figure \ref{fig:ddns} we compares the vertical structures of the 
nondimensional density perturbations, which are calculated as
\begin{equation}
\langle \frac{\delta \rho}{\rho}\rangle_{\Delta t_{\rm ave},x,y} 
= \frac{\langle\sqrt{(\rho - \langle \rho \rangle_{x,y})^2}
\rangle_{\Delta t_{\rm ave},x,y}}{\langle\rho\rangle_{\Delta t_{\rm ave},x,y}}. 
\end{equation}
The cases with $\gamma = 1.03$ (dashed line) and 1.0 (dotted line) exhibit 
similar vertical structures; larger $\delta \rho / \rho\approx 0.5$ in 
the surface regions decrease to $\sim 0.05$ at the midplane. 
On the other hand, the case with $\gamma=1.4$ shows complicated structure. 
The jumps at $\int_{0}^{z}dz/\langle h(z)\rangle_{\Delta t_{\rm ave},x,y}
\approx \pm 1.5$ coincide with the transition 
regions between the upper hot coronae and the lower cool midplane. 
In the midplane region, $\delta \rho / \rho\approx 0.1$ is larger than those 
obtained in the other two cases, probably because of the confinement of the 
sound-like waves in the midplane region (Figure \ref{fig:wave}).
However in the coronal regions $\delta \rho / \rho \lesssim 0.2$ 
is much smaller than those in the other cases.  

\subsection{Time Evolution of $B_y$}
\label{sec:By}

\begin{figure}[h]
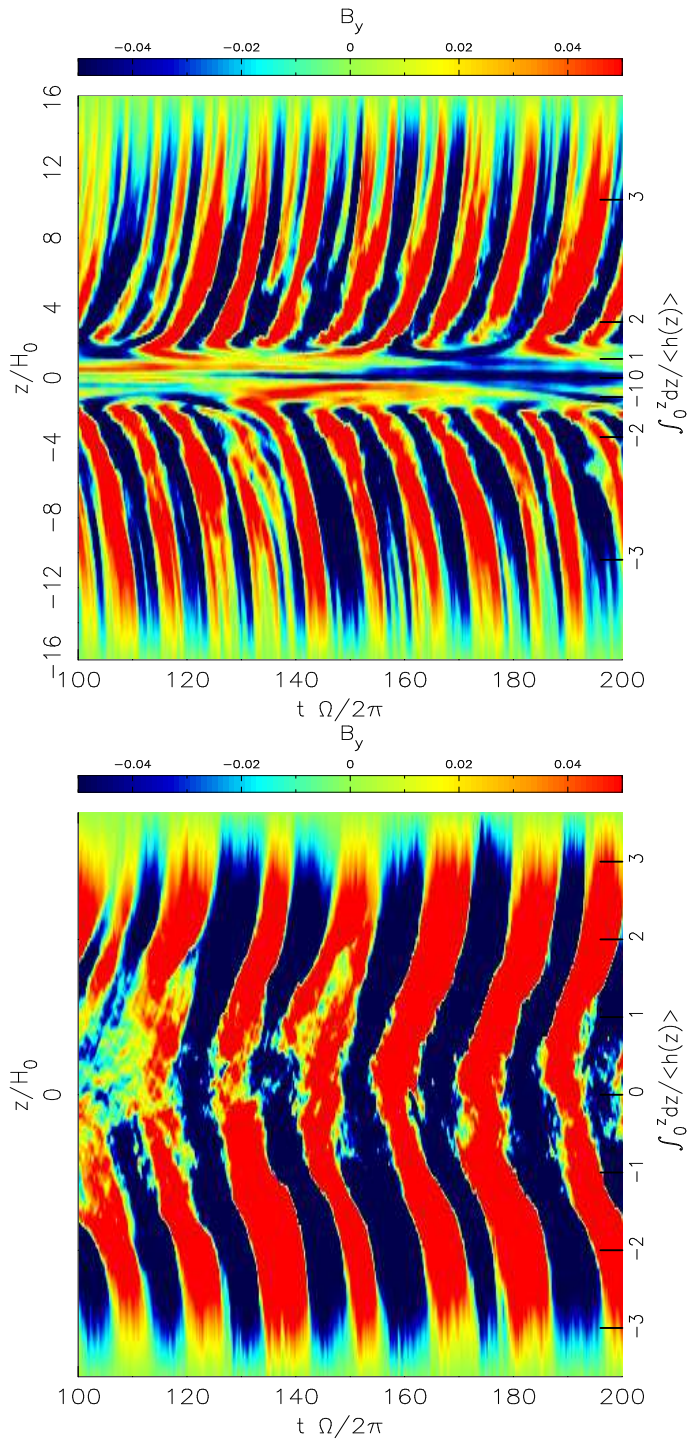

\begin{center}
\includegraphics[width=0.5\textwidth,height=0.4\textheight]{t-z_gamma14_By1.ps}\\
\includegraphics[width=0.5\textwidth,height=0.4\textheight]{t-z_gamma10_By1.ps}
\end{center}
\caption{Same as Figure \ref{fig:tzrhovz} but for the toroidal magnetic 
fields.}
\label{fig:tzBy}
\end{figure}

The evolution of toroidal magnetic fields, $B_y$, is widely discussed 
in terms of ``dynamo'' activity in disks by MRI 
\citep{bra95,nis06,skh10}. In Figure \ref{fig:tzBy}, we compare 
the $t-z$ diagrams of the cases with $\gamma=1.4$ (upper panel) and 
1.0 (lower panel). The isothermal case (lower panel) exhibits usual 
quasi-periodic changing of the sign of $B_y$, as seen by previous works 
\citep[e.g.][]{dav10}. 
The case with $\gamma=1.4$ also shows quasi-periodic oscillations of the 
$B_y$, but the period is shorter than that in the isothermal case and 
the amplitude (contrast between red and blue regions) is smaller. 
The tendency is similar for other large $\gamma(\ge 1.1)$ cases.
This indicates that the toroidal magnetic fields 
change the sign before the magnetic fields are amplified to the level as 
strong as that obtained in the isothermal case. This is probably because 
in the large $\gamma$ case the magnetic fields are more subject to the gas 
motion because of the high $\beta$ condition (Figure \ref{fig:vst_zHt_ab}).

The difference of the time-dependencies is consistent with the time 
evolutions of $\langle \alpha \rangle_{x,y,z}$ as shown in 
Figure \ref{fig:totalp}.
For larger $\gamma$, $\langle \alpha \rangle_{x,y,z}$ shows smaller 
fluctuations with time, 
because the dynamics are largely controlled by the gas pressures, which 
distribute more uniformly in a more time-steady manner, rather than by the 
magnetic fields, which intermittently pile up associated with channel flows.

Another interesting feature of the non-isothermal case with $\gamma=1.4$ 
is that the butterfly pattern triggered in the coronal regions is very 
vague in the midplane region. The left panel of Figure \ref{fig:efx} also 
shows that the Poynting flux is negligibly small in this region. They 
imply that the miplane region seems to be protected from the magnetic 
perturbations in the coronal regions. The transition region which separates 
the cool midplane and the hot corona accompanies the large density difference
to satisfy the pressure balance structure (Figure \ref{fig:vst_zH0}).  
This inevitably leads to the large jump of the \Alfven speed, $v_{{\rm A},z}$, 
across the transition region. Then, \Alfvenic perturbations arising from 
magnetic tension suffer reflection and the downward Poynting flux from the 
coronal regions are mostly reflected back and cannot penetrate the transition 
regions to the midplane. 
Reflection of \Alfven waves is widely discussed on the sun because it is 
very efficient at the transition region between the cool chromosphere and 
the hot corona \citep[e.g.,][]{suz05,ms12}. 
A signature of reflected \Alfven waves is actually observed 
on the solar surface by the HINODE satellite \citep{ft09}. 
These are quite similar to what we observe in the present simulations 
with relatively large $\gamma$.

\section{Summary and Discussion}
Using the 3D MHD simulations in the local stratified shearing boxes with 
weak net vertical magnetic fields, we have studied the formation of 
the hot coronae and the disk winds 
induced by the MRI turbulence in the accretion disks.  
Taking into account the effect of the cooling phenomenologically with
the effective ratio of the specific heats, $\gamma$, 
we have inspected how the basic properties are affected by the cooling.
Since the simulations are performed in the nondimensional form without 
any physical scale, the simulations are applicable to various objects. 
The amplifications of the magnetic fields are observed in all 
the simulation runs with different $\gamma$ from 1 to 5/3, which give the 
time- and box-averaged $\alpha\approx (1-3)\times 10^{-2}$, whereas 
we should take these values with cares because the numerical resolution is 
not sufficient in the midplane region. 

The properties of the coronae and the disk winds are not so affected by 
the numerical resolution because the simulations well capture the MRI there, 
and we have found that the results are classified into the two regimes.
In the small $\gamma(<1.03)$ regime, the temperatures in the surface regions 
are not high because the effect of the heating 
is weak owing to the small $\gamma$. 
The vertical outflows are directly driven by the Poynting flux 
associated with the amplified turbulent magnetic fields, they are more 
time-dependent, involved with the intermittent breakups of large-scale 
channel flows. 

In the large $\gamma(\ge 1.1)$ regime, the hot coronae form by the dissipation 
of the magnetic energy and the temperatures are $\sim 50$ times of the initial 
values with the time-averaged peak temperature slowly increasing 
with $\gamma$. 
The vertical outflows are mainly driven by the gas pressure of the hot coronae. 
Because the spatial distribution of the gas pressure is more uniform than 
that of the magnetic energy, the disk winds stream out in a more time-steady 
manner than in the small $\gamma$ regime. 
The hot coronae are connected to the cool midplane through the sharp 
transition regions.  Across the transition regions, both the sound and 
\Alfven speeds change abruptly because of the jumps of the 
temperature and the density. The transition regions work as the walls 
against wave activity. Sound-like waves are confined in the cool 
midplane region with giving the larger amplitudes of the density 
perturbations. 
The midplane region is also protected from the magnetic perturbations in 
the upper coronae.  

Although the driving mechanisms and the time-dependencies of the vertical 
outflows are different for the small and large $\gamma$ regimes, 
the time-averaged nondimensional mass fluxes, $C_{\rm W}$, are similar 
each other. 
This is because in both the regimes the origin that drives the vertical 
outflows is the magnetic energy that is amplified by the MRI. 
The magnetic energy is gradually converted at the locations with 
$\rho\approx 10^{-2}-10^{-3}\rho_{\rm mid}$,  
directly to the kinetic energy of the disk winds in the 
small $\gamma$ regime, or firstly to the thermal energy of the hot coronae 
that is finally transferred to the disk winds in the large $\gamma$ regime. 
The location of the energy conversion, which is
insensitive to $\gamma$, controls the final $C_{\rm W}$. 
We should cautiously note that the derived mass flux, 
$C_{\rm W}\approx 2\times 10^{-4}$, depends on the vertical box size 
\citep{suz10,fro13}, although the comparisons of different cases with 
the same vertical box size make sense. 

Our treatment of spatially uniform $\gamma$ is too much simplified. 
In reality, $\gamma$ should be non-uniform. Deeper regions near the midplane 
tend to be more optically thick, which gives larger $\gamma$, while in surface 
regions $\gamma$ is smaller.  
In more elaborated models with applications to specific objects, adequate 
cooling and heating processes should be included with radiative transfer 
\citep[e.g.,][]{hir06}. 
 
The authors thank Prof. Shu-ichiro Inutsuka for many fruitful discussion. 
The authors also appreciate constructive comments by an anonymous referee.
This work was supported in part by Grants-in-Aid for 
Scientific Research from the MEXT of Japan, 22864006.
Numerical simulations in this work were carried out with SR16000 at the Yukawa 
Institute Computer Facility and with the Cray XT4 and XC30 
operated in CfCA, National Astrophysical Observatory of Japan.


\end{CJK*}

\end{document}